\documentstyle[prl,aps,preprint]{revtex}
\begin{document} 
\draft
\tighten 
\title{Electron-photon interaction in resonant tunneling diodes}
\author{Jes\'us I\~{n}arrea, Ram\'on Aguado and Gloria Platero \\}
\address{Instituto de Ciencia de Materiales (CSIC),
Cantoblanco, 28049 Madrid, Spain.}
\maketitle   
\abstract{
We develope a model to describe the transmission coefficient  and
tunneling current
in the presence of
photon-electron coupling in a resonant 
diode. Our model takes into account multiphoton processes
as well as the transitions between
electronic states with different wave numbers.
This is crutial to explain the
experimental features observed in the tunneling current through
a double barrier which cannot be reproduced with more simplified
established models.
 According to our results, what experiments show in the current 
 density are quantum
photon-assisted features coming from multiphoton transitions 
which are not related with sample heating.}
\narrowtext
%twocolumn
Quantum transport through resonant heterostructures 
has been a very active       
research field in recent years, mainly due to the potential applicability
of such devices, specially from a technological point of view. 
When an external electromagnetic field is connected to these
semiconductor heterostructures, the physics interest and applications increase
dramatically, for instance 
we can cite the case of photocounters, light
detectors, microwave generators etc,. In the last years a great effort has
been done in theory\cite{1,2,3,4,5} an experiments \cite{6,7,8}
on transport in an 
AC field coupled to superlattices , double quantum wells
and quantum dots. However much less work has 
been devoted to the case of an electromagnetic field irradiating a
double barrier (DB)   
, not only from a theoretical point of view 
but also from
the experimental one \cite{2,9,10,11,12,13}. This last case, where
the coupling to the continuum of states in the
leads plays a crutial role in the current,
has to be analized including the transitions between different wave
number states
caused by the electron-photon interaction\cite{2,10,11}. 
The most popular model for photoassisted tunneling (PAT)
developed by Tien and Gordon (TG) \cite{14} 
assumes that the interaction with
the electromagnetic field is too weak to produce these transitions,
i.e., the electron-photon interaction is treated as an efective potential
$ V_{i} $ within every spatial region i of the system.
However, the exact coupling term in the Hamiltonian
 A.P ( A is the vector potential of the electromagnetic field
 and P the electronic momentum operator) produces transitions between 
different electronic states in the leads as well as in the quantum well.
In DB's the coupling of the well states with the continuum at
the 
leads produces a continuum of states in the well and the AC
field induces transitions between them which modify the transmission
coefficient and the tunneling current through the diode\cite{2,10}.
Then a description in terms of the TG model
which does not include those transitions is not suitable for describing
PAT through a DB. However, in systems like multiple quantum
wells, where the interwell sequential tunneling through quasidiscrete
states determines the tunneling current, the TG model can be
applied in most of the cases.\\ 
In this paper we calculate the coherent transmission coefficient
and the current through a DB under the      
presence of light linearly polarized in the growth direction. 
For that purpose we develope a 
quantum mechanical formalism based in  
a canonical transformation and in time-dependent perturbation
theory including multiphoton transitions.
In the qualitative behaviour observed experimentally
in the current density there are some features which at first sight 
do not seem to correspond to quantum PAT :
 the current threshold does not move linearly with the
field frequency, as expected in quantum photoassisted tunneling.
Also in the current density difference between the case with
and without radiation as a function of the external bias, a peak 
shows up at a fixed bias, independent of the field 
frequency\cite{13}.
The present
formalism gives good agreement with the
experiments\cite{13} and explains the observed features in the current                 
in terms of quantum response, excluding a trivial heating
effect from the laser pulse as responsible for them as well as
classical rectification.\\ 
The expression for the total Hamiltonian of the tunneling electron
under the influence of a laser polarized in the growth 
direction z is given by:
\begin{eqnarray}
H_{tot}=H_{e}^{0}+H_{ph}^{0}+W_{D}(t)+W_{OD}(t)
\end{eqnarray}
where
\begin{eqnarray}
H_{e}^{0}&=&\sum_{k} \epsilon_{k} c_{k}^{+} c_{k} \\
H_{ph}^{0}&=&\hbar w a^{+}a\\
W_{D}(t)&=&\sum_{k} [(e/m^{*})<k|P_{z}|k> c_{k}^{+} c_{k}
  (\hbar/2\epsilon Vw)^{1/2} (a e^{-iwt} +a^{+} e^{iwt})]\\
W_{OD}(t)&=& \sum_{k}\sum_{k^{'}\neq k} [(e/m^{*})<k^{'}|P_{z}|k>
  c_{k^{'}}^{+} c_{k} (\hbar/2\epsilon Vw)^{1/2}
   (a e^{-iwt} + a^{+} e^{iwt})]
\end{eqnarray}
being $w$ the photon
frequency. $H_{e}^{0}$ is the independent, 
electronic Hamiltonian and includes the double barrier
potential and the external applied bias, therefore 
the eigenstates of $H_{e}^{0}$, $\Psi_{0}(k)$, are the tunneling states 
for bare electrons.
$H_{ph}^{0}$, is the photon field Hamiltonian
and $W_{D}$ and $W_{OD}$, describe the coupling 
between electrons
and photons in the total Hamiltonian. 
Therefore the total Hamiltonian can be written as:
\begin{eqnarray}
H_{tot}=H_{D}(t)+W_{OD}(t)
\end{eqnarray}
where $H_{D}(t)=H_{e}^{0}+H_{ph}^{0}+W_{D}(t) $.
The hamiltonian $H_{D}$, can be solved exactly
considering a canonical transformation $ U=e^{s} $
where $ s= (e/m^{*}\hbar w) (\hbar/2\epsilon    
Vw)^{1/2} <k|P_{z}|k> c^{+}_{k} c_{k} (a^{+}
e^{i\omega t}-a e^{-i\omega t})$, obtaining an expression   
for the wave function of $H_{D}$:
\begin{eqnarray}
\Psi_{D}(k)=\Psi_{0}(k) \sum_{n=-\infty}^{\infty}J_{n}(\beta_{k}) e^{-inwt}
\end{eqnarray}
F and $ \omega$ are the electric field intensity and frequency 
of the laser field respectively, and  
$\beta_{k}=\frac{eF<P_{z}>}{m^{*}\hbar w^{2}}$.   
The dependence of the argument of the Bessel functions on the
inverse of $ \omega ^{2} $ instead on the inverse of the frequency
(as in TG model) comes from considering
the quantum mechanical hamiltonian with the electron-photon
coupling as in formula (4) and (5) (it can also be expressed in terms
of the position operator). This dependence was already
 obtained in ref.10 (see also ref.2 )
and recently it was discussed in terms of scaling\cite{12}.
Once we have obtained the eigenstate for $H_{D}$, we apply time-dependent
perturbation theory in order to treat the  $W_{OD}(t)$ term.
By doing this we have achieved an expression for the
total wave function of the tunneling electron under the influence
of a laser:
\begin{eqnarray}
\Psi(t)=\alpha[\Psi_{D}(k_{0})+\sum_{m}b_{m}^{(1)}(t)\Psi_{D}(k_{m})]
\end{eqnarray}
denoting by $k_{0}$, the wave vector of the initial electron, and $k_{m}$
the wave vector of the corresponding electronic coupled states. 
The coefficients $ b_{m}$ are given by:
\begin{eqnarray}
b_{m}^{(1)}=\frac{-ieFL}{4 \hbar^{2}w} \sum_{n^{'},n}
\left[ J_{n^{'}}(\beta_{k_{m}})J_{n}(\beta_{k_{0}}) 
\frac{<k_{m}|P_{z}|k_{0}>}{k_{m}} \right]
\end{eqnarray}
$n^{'}$ and $n$ run from $-\infty$ to $\infty$ and $m=n^{'}-n\pm1 =\pm1,
\pm2,\pm3,....$.
The normalization constant 
 $ \alpha=\frac{1}{\sqrt{1+\sum_{m}|b_{m}^{(1)}(t)|^{2}}}$, 
guarantees current conservation.
$\Psi_{D}(k_{0})$ is the "dressed" or diagonal 
reference state and
 $\Psi_{D}(k_{m})$, represents the coupled "dressed"
states due to photon absorption and emission.
The spectral density associated to (8) consists in a central peak
 (weighted
by $J_{0}^{2}$) and infinite n-sidebands separated in $n \hbar \omega $.
from the central peak and weighted by $ J_{n} $.
If the argument of the Bessel functions is very small, the sidebands
intensities are negligible and it is enough to consider 
transitions between the main side bands (the
ones weighted by $ J_{0}$) of different electronic states separated
in energy by $\hbar \omega$.
For higher values of the ratio: F/$ \omega^{2}$,
the spectral density weight is shared between the satellite peaks and
their contribution cannot be neglected.
According to our formalism, 
the electronic eigenvalues are shifted in
$\Delta=\frac{M^{2}}{\hbar w}$ being  
$M= \frac{e <P_{z}> }{m^{*}}(\frac{\hbar}{2\epsilon Vw})^{1/2}$. 
This shift in energy which can be expressed in terms of
F:  $ M=\frac{e<P_{z}>}{m^{*}}
\frac{F}{2 \sqrt{N} w}$ being $N$ the number of photons in the volume $V$,
results to be negligible 
with respect to the electron eigenvalues \cite{10,15}. 
Applying the current operator to the transmitted and incident
wave function, we obtain the time-averaged coherent transmission
coefficient following the Transfer Matrix technique:
\begin{eqnarray}
T&=&\frac{T_{0}}{(1+k_{1}/k_{0}|b_{1}^{(1)}|^{2}+k_{-1}/k_{0}
|b_{-1}^{(1)}|^{2}+k_{2}/k_{0}|b_{2}^{(1)}|^{2}+
k_{-2}/k_{0}|b_{-2}^{(1)}|^{2}+...... )}+\nonumber\\
 & &\frac{T_{1}|b_{1}^{(1)}|^{2}}{(k_{0}/k_{1}+|b_{1}^{(1)}|^{2}+k_{-1}/k_{1}
 |b_{-1}^{(1)}|^{2}+k_{2}/k_{1}|b_{2}^{(1)}|^{2}+ 
k_{-2}/k_{1}|b_{-2}^{(1)}|^{2}+...... )}+\nonumber\\
 & &\frac{T_{-1}|b_{-1}^{(1)}|^{2}}{(k_{0}/k_{-1}+k_{1}/k_{-1}
|b_{1}^{(1)}|^{2}+ |b_{-1}^{(1)}|^{2}+k_{2}/k_{-1}|b_{2}^{(1)}|^{2}+ 
k_{-2}/k_{-1}|b_{-2}^{(1)}|^{2}+...... )}+\nonumber\\
 & &\frac{T_{2}|b_{2}^{(1)}|^{2}}{(k_{0}/k_{2}+k_{1}/k_{2}|b_{1}^{(1)}|^{2}+
 k_{-1}/k_{2}|b_{-1}^{(1)}|^{2}+|b_{2}^{(1)}|^{2}+
 k_{-2}/k_{2}|b_{-2}^{(1)}|^{2}+...... )}+\nonumber\\
 & &\frac{T_{-2}|b_{-2}^{(1)}|^{2}}{(k_{0}/k_{-2}+
 k_{1}/k_{-2}|b_{1}^{(1)}|^{2}+
 k_{-1}/k_{-2}|b_{-1}^{(1)}|^{2}+k_{2}/k_{-2}|b_{2}^{(1)}|^{2}+
 |b_{-2}^{(1)}|^{2}+...... )}+\nonumber\\
 & &+......
\end{eqnarray}
The results for the transmission coefficient (see figure 1)
has been obtained for a
 $ Ga_{x}Al_{1-x}As $ DB with well and barrier 
thicknesses of $ 50 \AA$, and for F=$4.10^{5}
V/m$ and $\hbar \omega=13.6 meV$. The carrier density is
$n=10^{18}cm^{-3}$.
The main features observed in the transmission coefficient, T(E), 
are multiple
satellite peaks at both sides of the central one, coming from  
photon absorption and emission. The two closest peaks to the central
one correspond to one photon processes,
mainly to the transitions between the zero-side
bands of electronic states differing in one photon energy .The
transitions between higher side bands of different states differing in
energy one photon have very low intensities 
for these parameters and give a very small contribution to 
these two peaks. The other two peaks separated 2$ \hbar \omega$ 
from the main one 
correspond to 
processes involving two photons. Higher
multiphoton transitions has a much weaker intensity.\\ 
In fig. 2(a)  
we have plotted J/V
in the presence of the FIR laser. 
The effect of the light on J
can be observed in figures 2(b) and 2(c) corresponding 
to $F=4\times 10^{5}$ V/m for both cases 
and $ \hbar\omega$=13.6 meV and $ \hbar \omega $=4.2 meV respectively.
In those figures we plot 
 the current difference $\Delta J$
between the case where the light is present
and the case where there is no light.          
One observes firstly that the current threshold takes
place at the same bias, 
a main peak as well takes place for both cases
 at the same bias which corresponds
to the bias for the current threshold without light.        
 A shoulder appears for V close to                             
the center 
of the current peak, a weak negative contribution shows up
for higher bias and finally a small structure appears which is     
associated with the current cutoff. The fact that the current
threshold takes place at lower bias than in the case without
radiation is easy to understand: the electrons in the emitter
absorb photons and the current flows for $ E_{r}$ (energy of the
resonant state in the well) separated $ n\hbar\omega$ from
the emitter Fermi energy.
Then, regarding superficially to PAT theory, one would
expect, at first sight,  
not only that the current threshold shift linearly with 
$ \omega$ but also a dependence of the position in bias 
of the main peak
in $ \Delta J$ with $\omega$. However the experiments do not
show this frequency dependence\cite{13}. Some attempts have been made
in order to explain those results, first in terms of classical
response
and secondly relating the
experiments to sample heating due to the laser.
The dependence of J with the temperature
has been measured and shows qualitatively different behaviour than
the obtained in the presence of the laser, therefore the heating does
not explain the cited experiments\cite{13}.
 Regarding the classical response, the photon
energies considered are much larger than the energy broadening of
the DB resonant state and a quantum behaviour is expected.\\
The agreement between the experimental PAT current
and our results allows us to explain the features
disscused above. The fact that the current threshold
for different $ \omega$ takes place at practically the same bias
is related to the multiphotonic processes.
Comparing in fig.2, the two cases corresponding to $\omega$=13.6 meV and
$\omega$= 4.2meV for fixed F, we observe that for the case of lower
energy, as F/$\omega^{2} $ is larger more side bands contribute efficiently
to the current and therefore the J threshold moves to
lower bias. In this case n side bands have a non-negligible contribution
to J such that nx4.2$\sim m$x13.6 where m is the
number of side bands which support the current for $\hbar
\omega$=13.6meV (m smaller than n). \\
These multiphoton contributions wash out the
linear dependence that the threshold bias should follow as a 
function of $\hbar \omega$ if only one photon process took place.
The explanation for the same position in the main peak in $\Delta J$
is more subtle and it is related with the number of parallel states
available to tunnel resonantly with the absorption of one or more
photons. The main peak corresponds to the bias where the resonance
energy, $ E_{r}$,
is just above the fermi energy, $ E_{F}$, at 
the emitter. In this situation the 
number
of parallel states which can tunnel resonantly via absorption of
one or more photons is maximum no matter the $ \omega$ we are considering.
When $ E_{r}$ crosses $ E_{F}$ the absorption proccesses are now
compensated by emission from the emitter states to $ E_{r}$
and this fact reduces dramatically the general efficiency of the
resonant assisted tunneling.\\
In conclusion, we explain for the first time 
all the features in the current
through a DB induced by an electromagnetic 
field\cite{13} by means of a theory which includes 
the mixing of electronic states with different wavevector 
by the external field and
multiphoton transitions. 
Previous models which do not consider the mixing
of states, are not suitable for explaining the PAT
through a double barrier\cite{14}. More elaborated models
including this mixing but considering single photon 
proccesses\cite{10} could not explain, in 
terms of PAT, the independence of
the current threshold or cut-off with $\omega$.\\ 
Our present model explains
the available data in terms of
quantum response, excluding classical response and laser heating.\\
\par
 This work has been supported by the Comision Interministerial
de Ciencia y Tecnologia of Spain under contract MAT 94-0982-c02-02.\\
%\nonum

\begin{figure}
\caption{$ Log_{10}$ of coherent T(E) through a DB (well and barriers 
$ 50 \AA $ wide).  
 ($F=4 \times 10^{5}$ V/m, $\hbar w=13.6$ meV).}   
\end{figure}
\begin{figure}
\caption{ a) Coherent current as a function of DC bias 
for a $ Ga_{.7}Al_{.3}As$ DB (well and barriers $ 50 \AA$ wide).
b) $ \Delta J$ /V 
 (F=$4x10^{5} V/m $, $ \hbar \omega=13.6 meV$).
c) Same as in b) for $\hbar \omega=4.2 meV$).}
\end{figure}
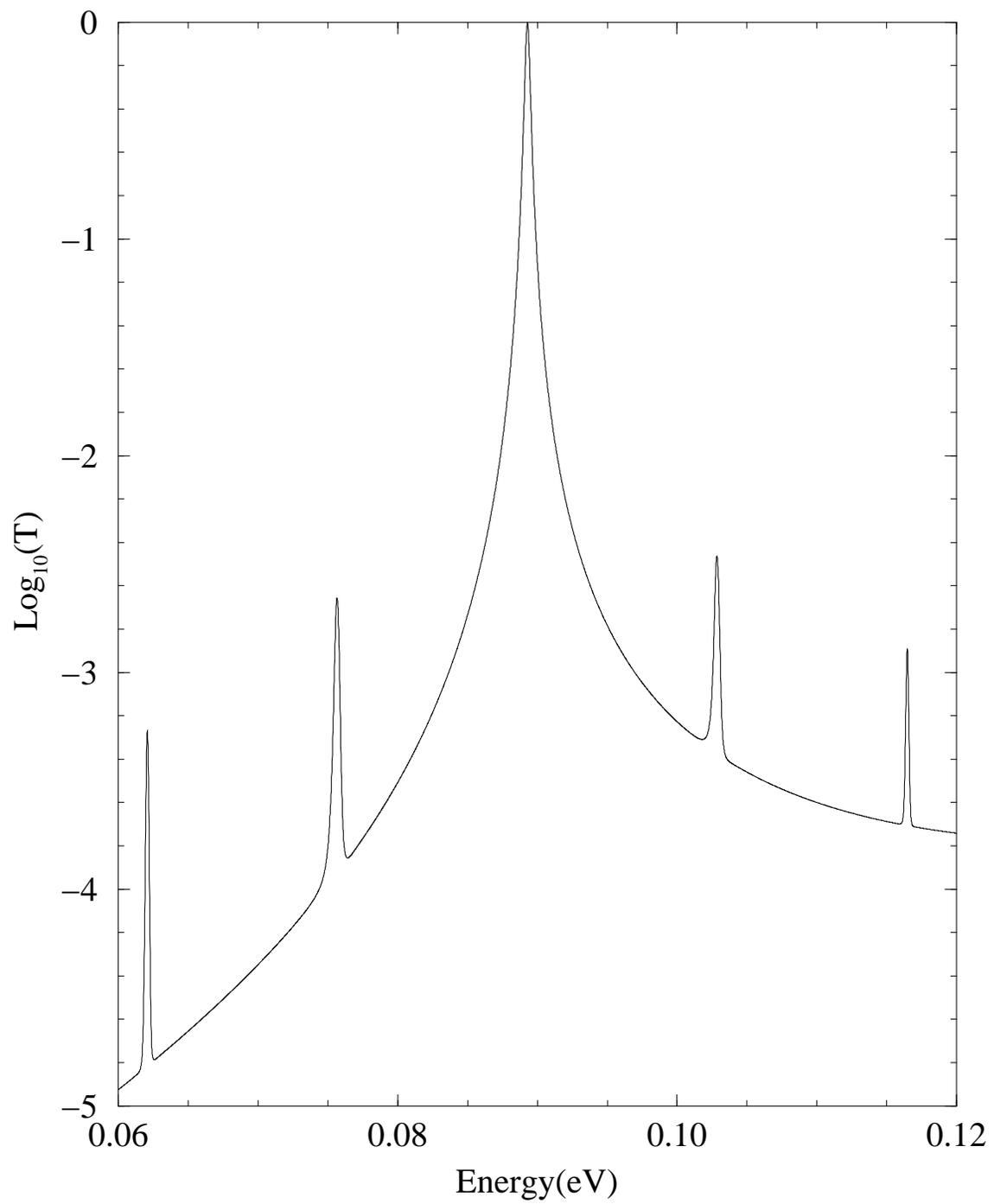
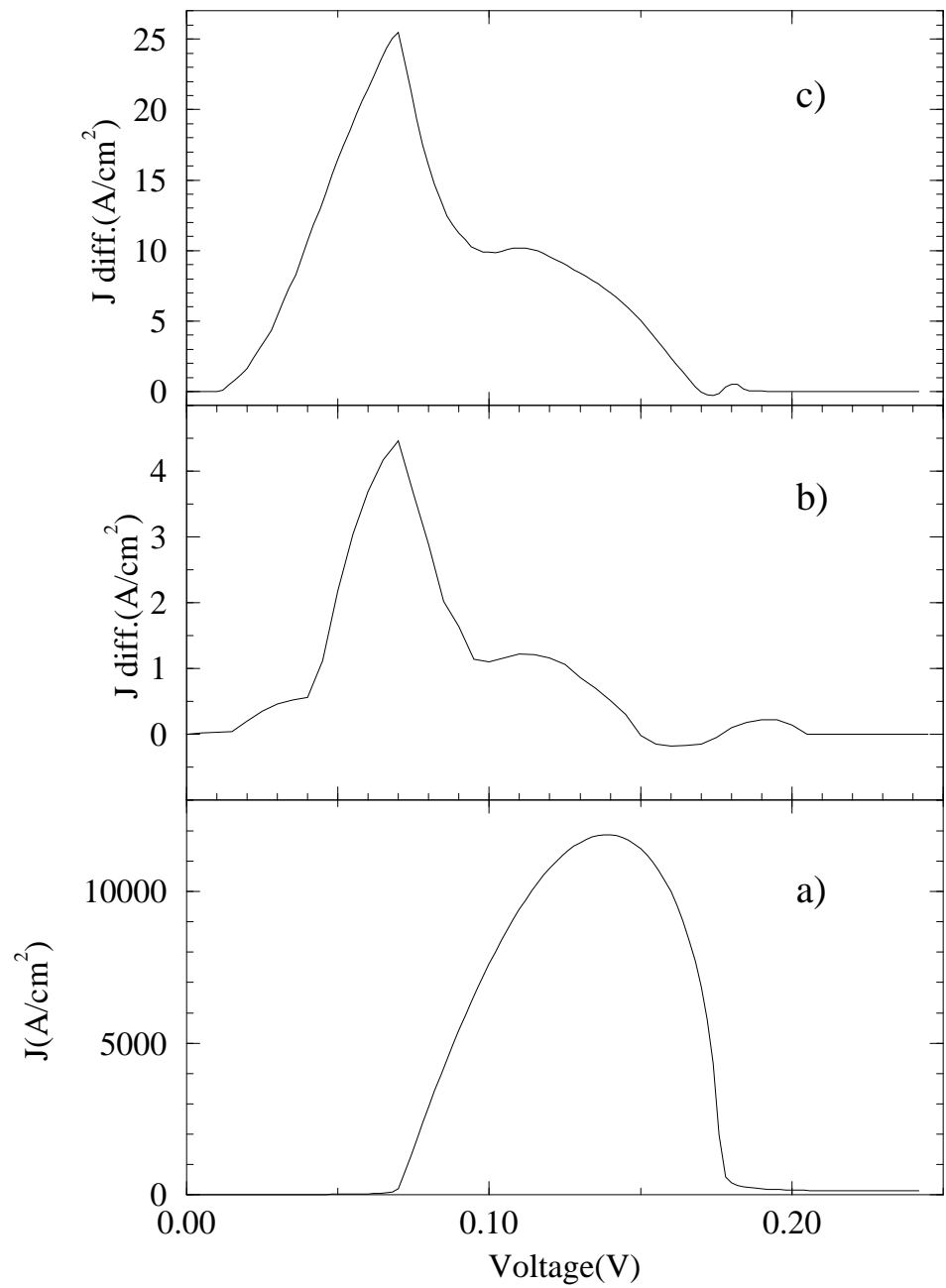
\end{document}